# Model-Driven GPR Inversion Network With Surrogate Forward Solver

Huilin Zhou, Xin Liu, Kexiang Wang and Shufan Hu

*Abstract*—Data-driven deep learning is considered a promising solution for ground-penetrating radar (GPR) full-waveform inversion (FWI), while its generalization ability is limited due to the heavy reliance on abundant labeled samples. In contrast, Deep unfolding network (DUN) usually exhibits better generalization by integrating model-driven and data-driven approaches, yet its application to GPR FWI remains challenging due to the high computational cost associated with forward simulations. In this paper, we integrate a deep learning-based (DL-based) forward solver within an unfolding framework to form a fully neural-network-based architecture, UA-Net, for GPR FWI. The forward solver rapidly predicts B-scans given permittivity and conductivity models and enables automatic differentiation to compute gradients for inversion. In the inversion stage, an optimization process based on the Alternating Direction Method of Multipliers (ADMM) is unfolded into a multi-stage network with three interconnected modules: data fitting, regularization, and multiplier update. Specifically, the regularization module is trained end-to-end for adaptive learning of sparse target features. Experimental results demonstrate that UA-Net outperforms classical FWI and data-driven methods in reconstruction accuracy. Moreover, by employing transfer learning to fine-tune the network, UA-Net can be effectively applied to field data and produce reliable results.

*Index Terms*—Deep learning, deep unfolding network (DUN), full-waveform inversion (FWI), model-driven.

## I. INTRODUCTION

GROUND-PENETRATING radar (GPR) is a near-surface geophysical technique widely used in various fields such as engineering investigation [1], agricultural soil studies [2], archaeology [3], and polar exploration [4]. This technique transmits high-frequency electromagnetic waves (from MHz to GHz) into the subsurface and thereby infer subsurface structures by analyzing the recorded signals [5]. Since the signals do not intuitively represent subsurface properties, processing is essential to clearly characterize subsurface anomalies [6], [7]. Among GPR imaging algorithms, full-waveform inversion (FWI) achieves high-resolution (half-wavelength) reconstruction of subsurface properties [8], [9]. However, due to the inherent nonlinearity and ill-posedness of the inverse problem, the success of FWI depends on an accurate initial model and proper regularization techniques [10], [11], [12], [13]. Moreover, FWI incurs prohibitive computational costs because forward simulation and gradient computation are required for each iteration. As a result, rapid and robust inversion of GPR data remains a significant challenge.

With advancements in deep neural network, several representative methods for reconstructing subsurface images from recorded signals have emerged, including PINet [14], GPRNet [15], GPRInvNet [16], GPRI2Net [17], DMRF-UNet [18] and 3DInvNet [19]. In recent years, many studies have employed diverse attention mechanisms to improve model perception of key features, enabling more accurate reconstruction of subsurface structures from complex GPR data [20], [21], [22], [23], [24]. These deep learning-based (DL-based) approaches learn the mapping relationship between GPR B-scans and permittivity models from massive labeled samples, which can be applied to new data for real-time inversion. However, these approaches are purely data-driven and lack physical interpretability, which leads to limited generalization performance in scenarios where sufficient labeled samples are unavailable.

To address this issue, researchers have attempted to incorporate domain knowledge into data-driven approaches. Huang et al. [25] proposed CL-CNN, which uses a closed-loop architecture to ensure that the output of the inversion network after a forward subnetwork remains consistent with the original B-scan image. Zhang et al. [26] constructed a cyclic generative network (GPR-CUNet) to achieve collaborative optimization of GPR forward and inversion processes, and they employed a hybrid loss function that incorporates predictive consistency and cyclic consistency as the overall learning strategy. Zhang et al. [27] proposed a deep learning inversion model based on a prior velocity field (VFIUNet), which integrates a multiscale transformation module (MSTM) and a hybrid loss function, significantly improving the reconstruction accuracy of permittivity distributions in subsurface root scenes. Huang et al. [28] pioneered the introduction of diffusion models into GPR inversion by proposing DGPRI-Net, which integrates UNet++, Visual Transformer (ViT), and the simple parameter-free attention (SimAM) module. In addition, Sun [29] and Wang [7] adopted migration results to provide correct time-depth relationships for deep learning inversion.

A further line of research in GPR data inversion focuses on model-based approaches enhanced by deep learning. A representative example of this category is the deep unfolding network (DUN), which has recently attracted attention due to its ability to provide accurate predictions with limited training samples. DUN originates from Plug-and-Play (PnP) methods

This work was supported in part by the National Natural Science Foundation of China under Grant 62161027. (Corresponding author: Shufan Hu.)

Huilin Zhou, Xin Liu and Kexiang Wang are with the School of Information Engineering, Nanchang University, Nanchang 330031, China (e-mail: zhouhuilin@ncu.edu.cn; 15763363067@163.com; 416100220004@email.ncu.edu.cn)

Shufan Hu is with the School of Mathematics and Computer Sciences, Nanchang University, Nanchang 330031, China (e-mail: shufanhu@ncu.edu.cn)

[30], [31], [32], [33], which use well-trained denoisers to implicitly represent the regularization term in optimization problems [34]. Inspired by PnP, DUN unfolds the iterative optimization algorithm into a multi-stage network, where each stage corresponds to one iteration of the optimization process. In each stage, the trainable denoisers are optimized together with other modules through backpropagation [35], [36]. Since the network implements the refinement of physical parameters rather than the mapping from signals to physical parameters, DUN is not restricted to fixed observation parameters, which theoretically tends to generalize better than data-driven methods. Up to now, DUN has been applied to various fields, including compressive sensing [37], [38], [39], medical imaging [40], [41], image restoration [42], and seismology [43], [44], demonstrating its potential in solving ill-posed inverse problems. Nevertheless, the additional computational cost of forward simulation and gradient computation still limits this approach from achieving real-time inversion.

By integrating a DL-based forward solver within an unfolding framework, this paper proposes a fully neural-network-based structure for GPR FWI, named UA-Net. Specifically, based on the Alternating Direction Method of Multipliers (ADMM), UA-Net decomposes the optimization problem into three unconstrained subproblems corresponding to data fitting, regularization, and multiplier updates. The solutions of these subproblems are then unfolded into a multi-stage network architecture, enabling the optimization process can be trained in an end-to-end manner. The DL-based forward solver is designed to rapidly generate GPR data predictions and accelerate the gradient computation in the data fitting module through automatic differentiation. The regularization module is designed as a soft-thresholding-based convolutional neural network (CNN), while the data fitting module and the multiplier update module are purely computational operations, with step sizes are automatically adjusted as learnable parameters.

The main contributions of this work are as follows:

(1) This work presents the first fully neural-network-based deep unfolding framework for GPR FWI, which enables differentiable, end-to-end training and thereby makes deep unfolding computationally viable for GPR FWI..

(2) The architecture, fine-tuned via transfer learning with limited samples, generalizes well to measured data, thus providing the possibility for rapid processing of field data.

The structure of this paper is as follows. Following the introduction, inverse problem formulation and ADMM-based solution are discussed. Then, the principles and architecture of the proposed UA-Net are introduced. Next, the effectiveness of the proposed method is verified through comparative experiments, and finally, the conclusions are presented.

## II. METHOD

In this section, we first introduce the formulation of GPR FWI. Then, we detail the inversion strategy based on ADMM. Finally, we introduce the proposed UA-Net.

### A. Inverse Problem Formulation

The relationship between GPR data and the physical properties of subsurface structures can be described as follows:

$$d_{obs} = F(m) + n \quad (1)$$

where $d_{obs}$ denotes the observed data; $F(\cdot)$ represents the forward modeling function; $m$ represents the model parameters, typically the permittivity and conductivity; and $n$ represents the measurement noise.

GPR FWI aims to infer model parameters $m$ from observed data $d_{obs}$. The inverse problem with regularization can be expressed as the following optimization problem:

$$\Phi(m) = \frac{1}{2} \| d_{obs} - F(m) \|_2^2 + \lambda R(m) \quad (2)$$

where $R(m)$ is the regularization term, and $\lambda$ is the regularization parameter.

### B. Inversion Strategy Based on ADMM

In this paper, the L1 regularization is adopted for considering the inherent sparsity of model properties. The objective function with L1 regularization can be expressed as follows:

$$\Phi(\varepsilon) = \frac{1}{2} \| d_{obs} - d(\varepsilon) \|_2^2 + \lambda \| D\varepsilon \|_1 \quad (3)$$

where $D$ is a transformation operator; $d(\varepsilon)$ denotes the modeled data, and the relative permittivity $\varepsilon$ is used for single-parameter inversion.

We decouple the data fitting term from the regularization term by introducing an auxiliary variable $z$. Equation (3) can be equivalently rewritten as follows:

$$\varepsilon^* = \arg\min_{\varepsilon} \left( \frac{1}{2} \| d_{obs} - d(\varepsilon) \|_2^2 + \lambda \| Dz \|_1 \right) \quad s.t. \, z = \varepsilon \quad (4)$$

Equation (4) can be reformulated as an unconstrained problem using the ADMM approach:

$$A_\gamma(\varepsilon, z, \alpha) = \frac{1}{2} \| d_{obs} - d(\varepsilon) \|_2^2 + \lambda \| Dz \|_1 \\ + \alpha^T(\varepsilon - z) + \frac{\gamma}{2} \| \varepsilon - z \|_2^2 \quad (5)$$

where $\alpha$ represents the Lagrangian multiplier and $\gamma$ is the Lagrangian parameter.

In this way, the solution to (5) can be obtained by alternately solving the following three subproblems:

$$\varepsilon_k = \arg\min_{\varepsilon} \frac{1}{2} \| d_{obs} - d(\varepsilon) \|_2^2 \\ + \frac{\gamma_k}{2} \| \varepsilon_{k-1} - z_{k-1} + \frac{\alpha_{k-1}}{\gamma_k} \|_2^2 \quad (6)$$

$$z_k = \arg\min_{z} \frac{\gamma_k}{2} \| \varepsilon_k + \frac{\alpha_{k-1}}{\gamma_k} - z_{k-1} \|_2^2 + \lambda_k \| Dz_{k-1} \|_1 \quad (7)$$

$$\alpha_k = \alpha_{k-1} + \omega_k(\varepsilon_k - z_k) \quad (8)$$

where $\omega$ is the update rate of the Lagrange multipliers, and $k$ is the number of iterations.

Equation (6) corresponds to the update of the permittivity model, which we solve using the classical gradient descent method:

$$\varepsilon_k = \varepsilon_{k-1} - \rho_k \left[ \nabla E_k + \gamma_k \left( \varepsilon_{k-1} - z_{k-1} + \frac{\alpha_{k-1}}{\gamma_k} \right) \right] \quad (9)$$

where $\nabla E$ is the gradient of the data fitting term, and $\rho$ represents the step size.

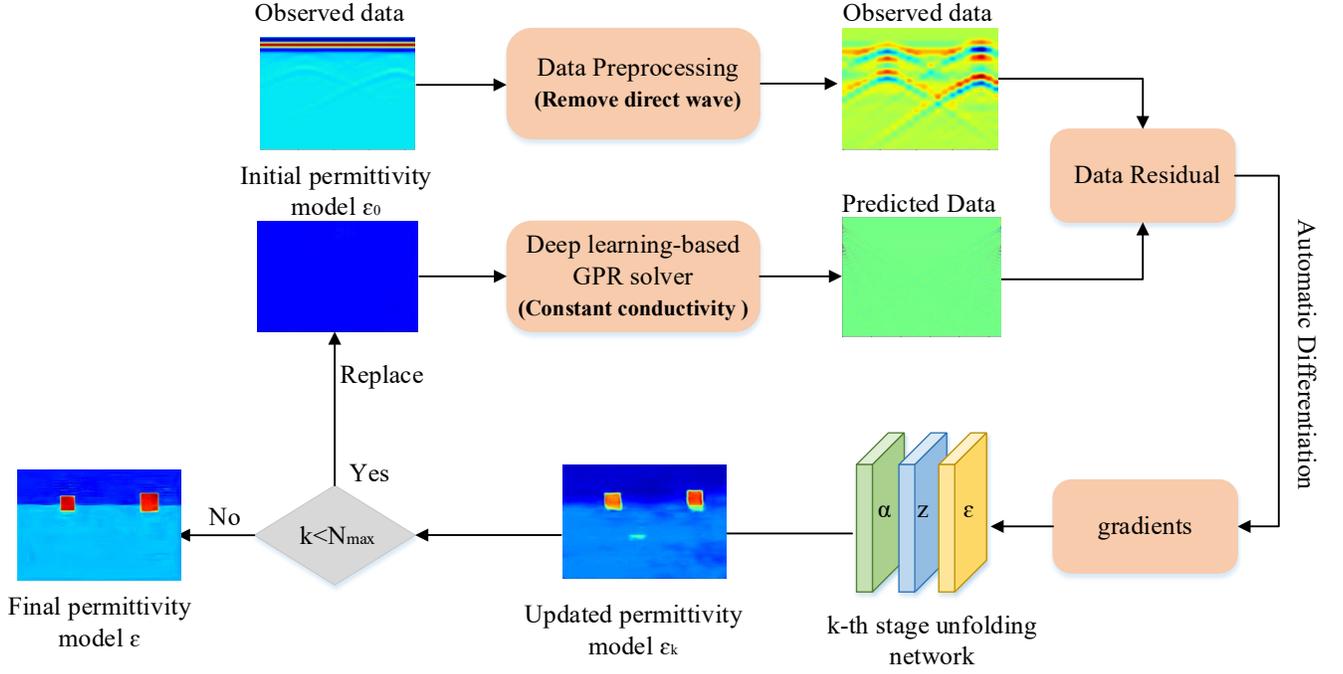

Fig. 1. Diagram of the workflow of UA-Net.

Equation (7) is actually a specific instance of proximal mapping, i.e. $\text{prox}_{\lambda\varphi}\left(\varepsilon_k + \alpha_{k-1}/\gamma_k\right)$, when $\varphi(z_k) = \|Dz_k\|_1$, and the proximal mapping of regularizer φ denoted by prox$_{\lambda\varphi}$(z) is defined as:

$$\text{prox}_{\lambda\varphi}\left(\varepsilon_k + \frac{\alpha_{k-1}}{\gamma_k}\right) = \arg\min_z \frac{\gamma_k}{2}\|\varepsilon_k + \frac{\alpha_{k-1}}{\gamma_k} - z\|_2^2 + \lambda_k\varphi(z_{k-1}) \quad (10)$$

Assuming D is an orthogonal matrix, the analytical solution of (7) can be expressed as follows:

$$z_k = D^T S_{\lambda_k/\gamma_k} D\left(\varepsilon_k + \frac{\alpha_{k-1}}{\gamma_k}\right) \quad (11)$$

where $S_{\lambda_k/\gamma_k}$ is a soft threshold function.

In this way, by iteratively solving (9), (11), and (8), the optimal estimation for the permittivity ε can be progressively approached

*C. Inversion Strategy Based on UA-Net*

As explained before, the GPR inversion problem can be solved using the ADMM-based iterative scheme. However, the design of the transformation matrix *D* and the selection of parameters could be challenge for achieving the desired result. In this section, we unfold the iteration of the three equations into a multi-stage network, where (11) is designed as a learnable regularizer based on a CNN, while (8) and (9) are purely computational operations, with the step size adjusting automatically as learnable parameters. For the gradient calculation in (9), we design a DL-based GPR solver that enables fast B-scan predictions and efficient gradient computation through automatic differentiation. We first give the overall workflow of UA-Net, and then detail the DL-based GPR solver and the multi-stage inversion network in sequence.

*1) Overall workflow of UA-Net*: We unfold (8), (9) and (11) into a multi-stage inversion network, where each stage corresponds to one iteration of the optimization process. The workflow is depicted in Fig. 1. First, the initial estimates of the permittivity and conductivity are input into the well-trained GPR solver to obtain B-scan predictions. Then, the gradient of the data fitting term is computed through automatic differentiation and fed into the inversion network. After that, the permittivity model *ε*, auxiliary variable *z*, and Lagrange multiplier *α* are updated in the inversion network. Notably, the permittivity model output from current stage serves as the initial permittivity model for the following stage, while the conductivity remains unchanged. The iteration process continues until the maximum iteration limit is reached.

*2) DL-based GPR solver*: We use an encoder-decoder neural network (U-Net) to form a mapping from relative permittivity *ε* and conductivity *σ* to the corresponding B-scan $\hat{y}$, which can be expressed as follows:

$$\hat{y} = h_\theta(\varepsilon, \sigma) \quad (12)$$

The inputs to the network are relative permittivity *ε* and conductivity *σ*, and the output is the predicted B-scan $\hat{y}$. A well-trained network enables fast prediction of B-scans based on the given model parameters. The encoder includes four encoding blocks, each consisting of two consecutive 3 × 3 convolutions, followed by a 2 × 2 max pooling for downsampling. A BatchNorm and a ReLU activation follow every convolution. Every encoding block doubles the feature channels. The decoder includes four decoding blocks, each consisting of upsampling, 2 × 2 convolutions to reduce feature channels, concatenation with corresponding encoder features, and two consecutive 3 × 3 convolutions. A BatchNorm and a ReLU activation also follow every convolution. A 1 × 1 convolution maps each 64-dimensional feature vector to a single one. In the final stage, the B-Scan of the specified dimensions is output through bilinear interpolation. The structure of the DL-based GPR solver is shown in Fig. 2.

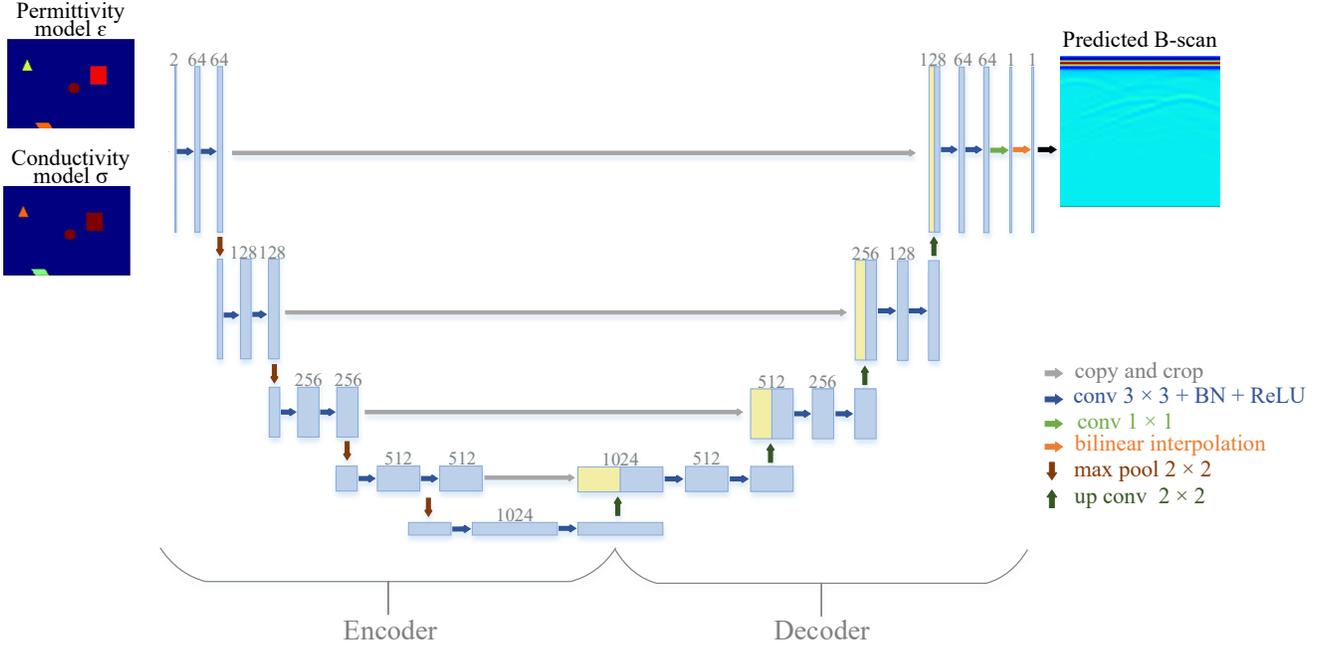

Fig. 2. Structure of the deep learning-based GPR solver.

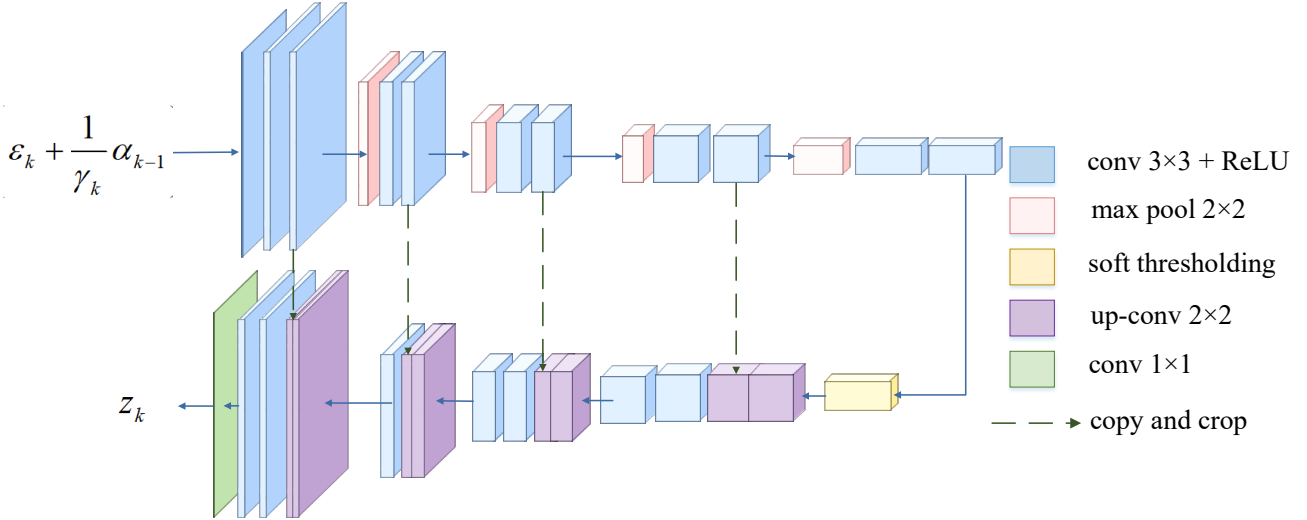

Fig. 3. Structure of the soft-thresholding-based learnable regularizer.

*3) Multi-stage inversion network*: Based on the detailed derivation in the Appendix, the proximal mapping (7) can be implemented by the following network:

$$z_k = G_k\left(soft\left(F_k\left(\varepsilon_k + \frac{1}{\gamma_k}\alpha_{k-1}\right), \theta_k\right)\right) \quad (13)$$

where $F(\cdot)$ and $G(\cdot)$ represent a symmetric network structure, and $\theta_k$ is a learnable shrinkage threshold.

We design a soft-thresholding-based U-Net to implement (13), where the encoder and decoder serve as symmetric components $F(\cdot)$ and $G(\cdot)$, respectively. As illustrated in Fig. 3, the network is composed of an encoder, a soft-thresholding module, and a decoder. The encoder contains four encoding blocks, each consisting of two consecutive 3 × 3 convolutions, followed by a 2 × 2 max pooling for downsampling, with a ReLU activation following every convolution. After the encoder, the extracted feature maps are processed by a soft-thresholding module, which applies adaptive shrinkage to enforce sparsity constraints. The decoder mirrors the encoder and including four decoding blocks. Each block begins with upsampling, followed by a 2 × 2 convolution. The upsampled features are then joined with the corresponding encoder features using skip connections. The features are subsequently passed through two 3 × 3 convolution layers, each followed by a ReLU activation. Finally, the network output is obtained through a 1 × 1 convolution.

In each stage, the regularization module is trained end-to-end along with learnable parameters in the data fitting module and the multiplier update module, while the parameters of the well-trained GPR solver remain unchanged. To increase network capacity, instead of sharing the same network parameters, each stage has its own $F(\cdot)$, $G(\cdot)$, and $\theta_k$. The specific detail of the three modules is as follows.

**Algorithm 1** UA-Net based Inversion Method
---
Inputs: Initial permittivity model $\varepsilon_0$; fixed conductivity model $\sigma$; auxiliary variable $z_0$ initialized to 0; Lagrange multipliers $\alpha_0$ initialized to 0; real B-Scan $d_{obs}$; maximum number of iterations $N_{max} = 5$.
Step 1: while $k \leq N_{max}$ do
Step 2:   Forward modeling: $\hat{d} = h_\theta(\varepsilon, \sigma)$
Step 3:   Gradient computation (automatic differentiation): $\nabla E_k$
Step 4:   Update permittivity model $\varepsilon$:
$$\varepsilon_k = \varepsilon_{k-1} - \rho_k \left[ \nabla E_k + \gamma_k \left( \varepsilon_{k-1} - z_{k-1} + \frac{1}{\gamma_k} \alpha_{k-1} \right) \right]$$
Step 5:   Update auxiliary variable $z$:
$$z_k = G_k \left( soft \left( F_k \left( \varepsilon_k + \frac{1}{\gamma_k} \alpha_{k-1} \right), \theta_k \right) \right)$$
Step 6:   Update Lagrange multiplier $\alpha$:
$$\alpha_k = \alpha_{k-1} + \omega(\varepsilon_k - z_k)$$
Step 7:   Set $k = k + 1$
Step 8: end while
Output: permittivity model $\varepsilon_k$

*a) Data fitting module $\varepsilon_k$:* This module updates the permittivity model according to (9). In the first stage ($k = 1$), the permittivity model is updated using the initial permittivity model $\varepsilon_0$, the gradient $\nabla E_0$ and the zero matrices $z_0$ and $\alpha_0$. In subsequent stages ($k > 1$), the permittivity model $\varepsilon_k$ is updated based on $\varepsilon_{k-1}$, $\nabla E_{k-1}$, $z_{k-1}$ and $\alpha_{k-1}$. The gradient in this module is computed using automatic differentiation.

*b) Regularization module $z_k$:* This module updates the auxiliary variable $z$ using a soft-thresholding based Convolutional Neural Network, which is shown in Fig. 3. In the first stage ($k = 1$), the permittivity model $\varepsilon_1$ and the zero matrix $\alpha_0$ are input into the network, and the updated $z_1$ is output. For subsequent stages ($k > 1$), the network takes the permittivity model $\varepsilon_k$ from the current stage and $\alpha_{k-1}$ from the previous stage as input, and outputs $z_k$.

*c) Multiplier update module $\alpha_k$:* The update for this module is defined in (8). In the first stage ($k = 1$), $\alpha_1$ is updated based on $\varepsilon_1$, $z_1$ and the zero matrix $\alpha_0$. For subsequent stages ($k > 1$), $\varepsilon_k$, $z_k$ from the current stage and $\alpha_{k-1}$ from the previous stage are used to update $\alpha_k$.

The pseudo-code of UA-Net is provided in Algorithm 1.

*4) Loss Function of UA-Net*: The expression of the loss function is:

$$Loss = \frac{1}{n} \sum_{k=1}^{n} \| \varepsilon_k - \varepsilon_t \|_2^2 \quad (14)$$

where $n$ is the total stages of the network, and $\varepsilon_t$ denotes the ground truth of the permittivity model.

## III. NUMERICAL RESULTS

To train and test the proposed UA-Net, permittivity and conductivity models are generated using MATLAB, and the corresponding GPR data are obtained using a 2D Finite Difference Time Domain (FDTD) algorithm [45]. The simulation area is $0.34 \times 0.52$ m² with gird size of $0.004 \times 0.004$ m². The time window is 20 ns and the source wavelet is Blackman-Harris pulse with a center frequency of 900 MHz. Both the transmitter and receiver are horizontal dipole antennas. The transmitter and receiver are situated at the same location and move together along a straight scanning path, with a scanning step of 2 cm. The subsurface objects have four shapes: rectangle, circle, triangle, and parallelogram. Each object has a random size, position, relative permittivity and conductivity. The relative permittivity and conductivity of the background model vary within [2, 6] and [2, 6] mS/m, respectively, while those of the objects are randomly selected from [8, 16] and [8, 16] mS/m, respectively.

All training and testing are performed on a PC equipped with an Intel(R) Core(TM) i7-10700 CPU and an NVIDIA GeForce RTX 3080 GPU, with the network implemented on Pytorch. In all experiments, the dataset is divided into 80% for training, 10% for validation, and 10% for testing. The model is evaluated on the validation set after each training epoch, and the final model is determined by its performance on the validation set.

To quantitatively evaluate the performance of the proposed method, we compare the peak signal-to-noise ratio (PSNR) and structural similarity index (SSIM) between the ground truths and the reconstructed permittivity models. The PSNR and SSIM are defined as follows:

$$PSNR = 20 \cdot \log_{10} \left( \frac{MAX_I}{\sqrt{MSE}} \right) \quad (15)$$

$$SSIM\left(\varepsilon_k^{rec}, \varepsilon_k^{gt}\right) = \frac{\left(2\mu_{\varepsilon_k^{rec}} \mu_{\varepsilon_k^{gt}} + C_1\right) \cdot \left(2\sigma_{\varepsilon_k^{rec} \varepsilon_k^{gt}} + C_2\right)}{\left(\mu_{\varepsilon_k^{rec}}^2 + \mu_{\varepsilon_k^{gt}}^2 + C_1\right) \cdot \left(\sigma_{\varepsilon_k^{rec}}^2 + \sigma_{\varepsilon_k^{gt}}^2 + C_2\right)} \quad (16)$$

where $\varepsilon_k^{rec}$ and $\varepsilon_k^{gt}$ represent the reconstructed and ground truths of permittivity models, respectively; $\mu_{\varepsilon_k^{rec}}$ and $\mu_{\varepsilon_k^{gt}}$ are the local means of the permittivity models; $\sigma_{\varepsilon_k^{rec}}$ and $\sigma_{\varepsilon_k^{gt}}$ are their corresponding standard deviations; $\sigma_{\varepsilon_k^{rec} \varepsilon_k^{gt}}$ denotes the covariance between $\varepsilon_k^{rec}$ and $\varepsilon_k^{gt}$; C1 and C2 are the regularization constants; and $MAX_I$ is the maximum pixel value of the image.

*A. Experiment I: Performance Evaluation of the DL-Based GPR Forward Solver*

For training the DL-based GPR solver, a dataset of 20 000 samples is utilized. The Adam optimizer is employed with an initial learning rate of 0.001, which is reduced by $1 \times 10^{-6}$ if the validation loss does not decrease after five epochs. Mean-squared error (MSE) is used as the loss function. Additionally, the batch size is set to 8, and the training is limited to a maximum of 200 epochs.

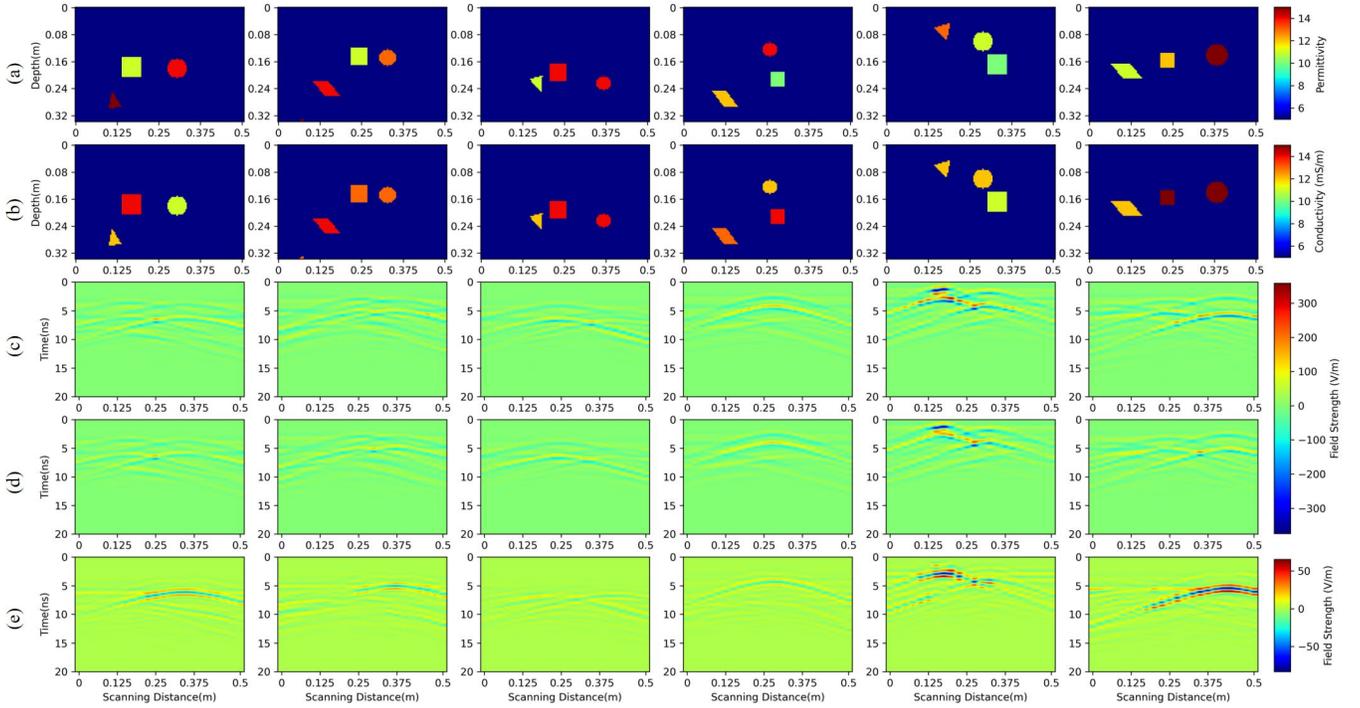

Fig. 4. Comparison of the forward simulation results. (a) Permittivity models; (b) conductivity models; (c) actual B-scans with direct wave removed; (d) predicted B-scans with direct wave removed; (e) absolute difference between actual B-scans and the predicted B-scans.

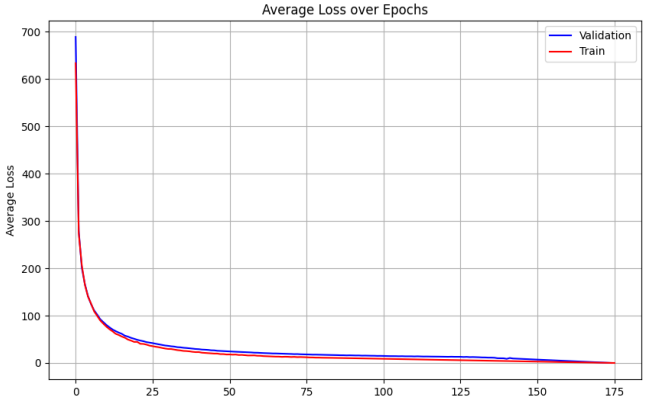

Fig. 5. Training and validation curves of the DL-based GPR solver.

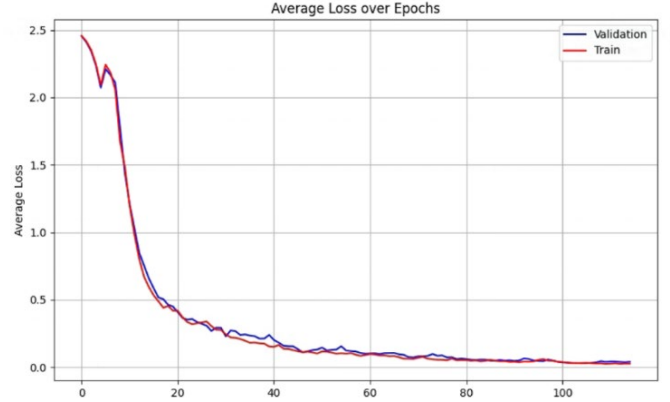

Fig. 6. Training and validation curves of UA-Net during inversion.

The training and validation loss curves of the GPR solver are shown in Fig. 5, indicating the convergence of the loss function. The B-scans predicted by the proposed DL-based GPR solver are compared with those generated by the FDTD solver. Fig. 4 present the comparative results for six models from the test dataset, with subsurface objects of various shapes, sizes, positions, permittivity, and conductivities. As shown in Fig. 4 (c) and (d), the GPR solver predictions match well with the actual B-scans, In Fig. 4 (e), we use absolute error to visually compare the predicted B-scan and the FDTD results. It can be seen that, although there is an energy difference on the strong-reflection hyperbolic wave, the shape of the hyperbolic wave is well preserved. Fig. 7 presents four cases that compare A-Scans from the DL-based GPR solver with ones obtained using the FDTD solver. For clearer observation of reflected waves, the direct waves have been removed from all four cases. It is obvious that the DL-based GPR solver predictions match well with ones obtained using the FDTD solver in both the waveform shape and the arrivals of the responses.

Quantitative validation was performed in TABLE I with the normalized root mean square error (NRMSE) consistently below 7.03%, demonstrating the high fidelity of the DL-based GPR solver in waveform reconstruction.

*B. Experiment II: Inversion Results and Comparative Study*

Building upon the accuracy of the proposed DL-based GPR forward solver demonstrated in Experiment I, we further evaluate the effectiveness of the proposed inversion framework. Specifically, we train UA-Net for permittivity reconstruction using a dataset of 1000 samples. The Adam optimizer is employed with an initial learning rate of 0.0001, which is reduced by $0.5 \times 10^{-6}$ if the validation loss does not decrease after five epochs. A batch size of 1 is used, and the training is limited to a maximum of 150 epochs. After repeated experiments and balancing computational cost with reconstruction accuracy, the maximum iteration count $N_{max}$ is set to 5. Fig. 6 shows the training and validation loss curves of UA-Net, indicating the convergence of the loss function.

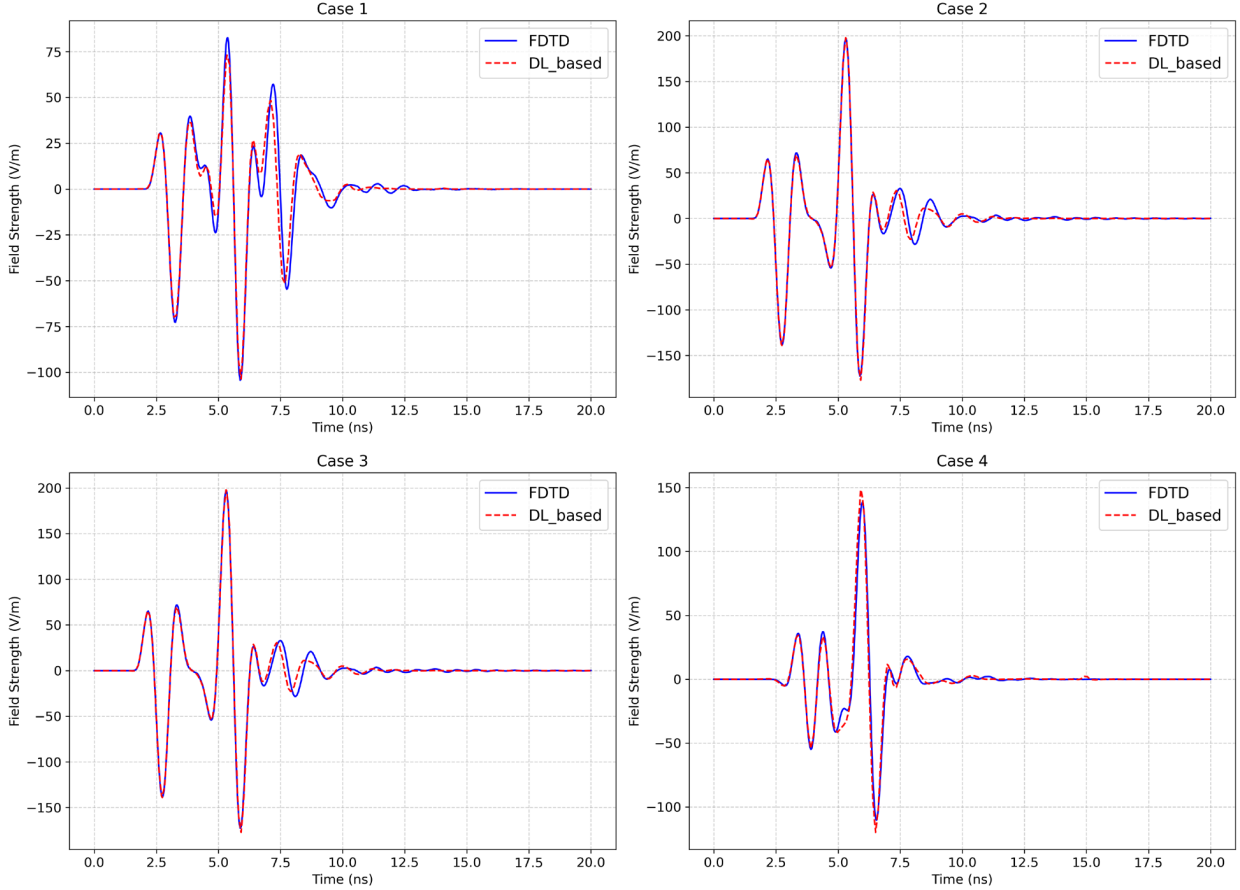

Fig. 7. Four different ascans used to compare the proposed DL-based GPR solver with the FDTD solver.

TABLE I
COMPARISON OF NRMSE FOR SIX TEST SAMPLES

| T1 | T2 | T3 | T4 | T5 | T6 |
|---|---|---|---|---|---|
| 4.76% | 4.85% | 2.90% | 4.12% | 3.26% | 7.03% |

We compare the permittivity reconstruction performance using three methods: classical FWI [46], PINet [14], and the proposed UA-Net. TV regularization is applied in classical FWI, with an initial regularization weight of 0.5. We employ a three-step multiscale FWI with center frequencies of 500, 700, and 900 MHz, with 50 iterations for each step. For PINet, we use a dataset containing 20 000 samples as it is purely data-driven. The training epochs for PINet are set to 200. Fig. 8 (a), (b), (c), and (d) present the ground truths of permittivity models and reconstructed permittivity distributions obtained using classical FWI, PINet, and UA-Net. It can be observed that classical FWI can only estimate the position and shape of the object but struggles to accurately reconstruct the permittivity. PINet exhibits distortion in different degrees. In Fig. 8 Cases 1, Case 2, and Case 3, the predictions of PINet show significant discrepancies in shapes and permittivity of triangular objects. In Fig. 8 Case 4, PINet fails to reconstruct the triangular object. In contrast, UA-Net provides good reconstruction performance for all four test samples, which also demonstrates that the proposed DL-based GPR solver is capable of effectively capturing the hyperbolic features of the object response in the B-scan and incorporating them into the inversion network.

Fig. 9 presents a comparative analysis of prediction accuracy among FWI, PINet, and U-Net using differential heatmaps. The results demonstrate that FWI exhibits high deviations in the reconstructed permittivity values of the object, while PINet exhibits noticeable object loss in its output, which may attributable to its limited capability in extracting critical features during the training phase. In contrast, UA-Net demonstrate the highest prediction consistency, with only minor deviations observed in marginal areas.

The performance of this method is quantitatively evaluated by computing the mean reconstruction time (MRT, excluding model loading), SSIM, and PSNR for the test samples, as detailed in TABLE II. Compared with classical FWI, the

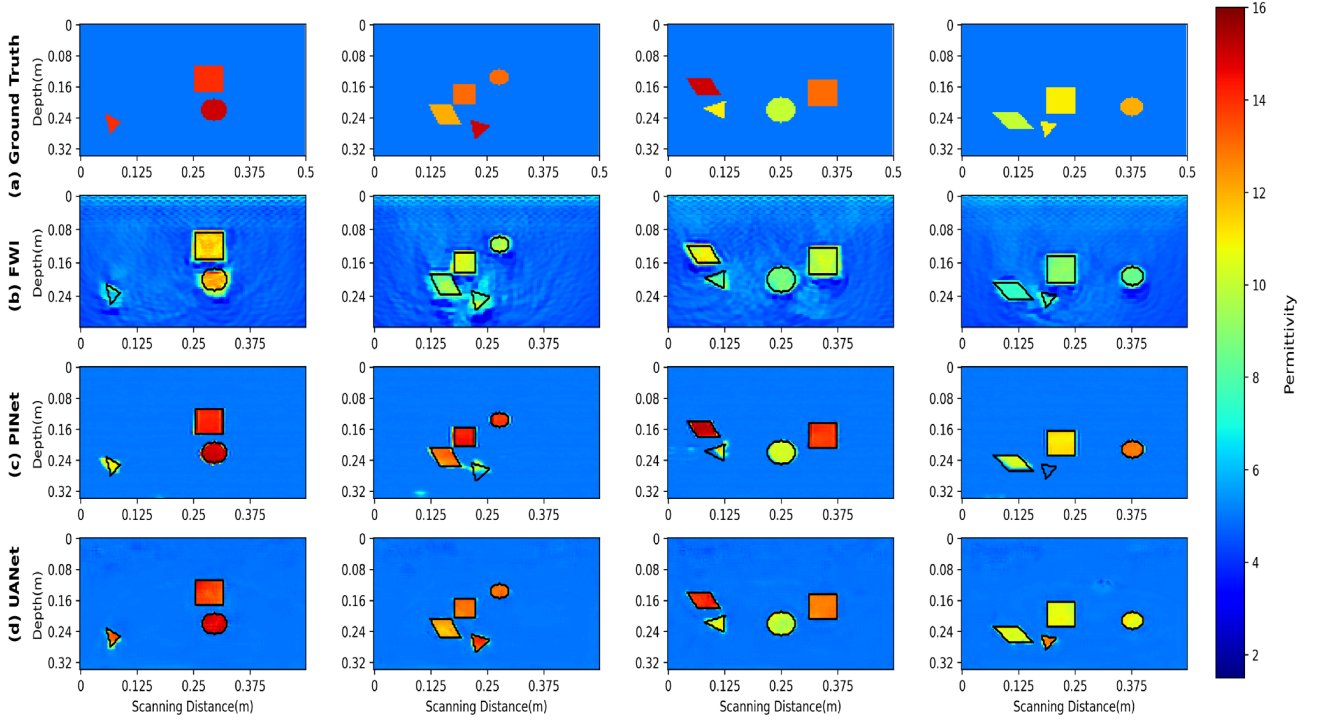

Fig. 8. Comparison of the inversion results. (a) Ground truths of the permittivity; (b) reconstructed permittivity models using classical FWI; (c) reconstructed permittivity models using PINet; (d) reconstructed permittivity models using UA-Net.

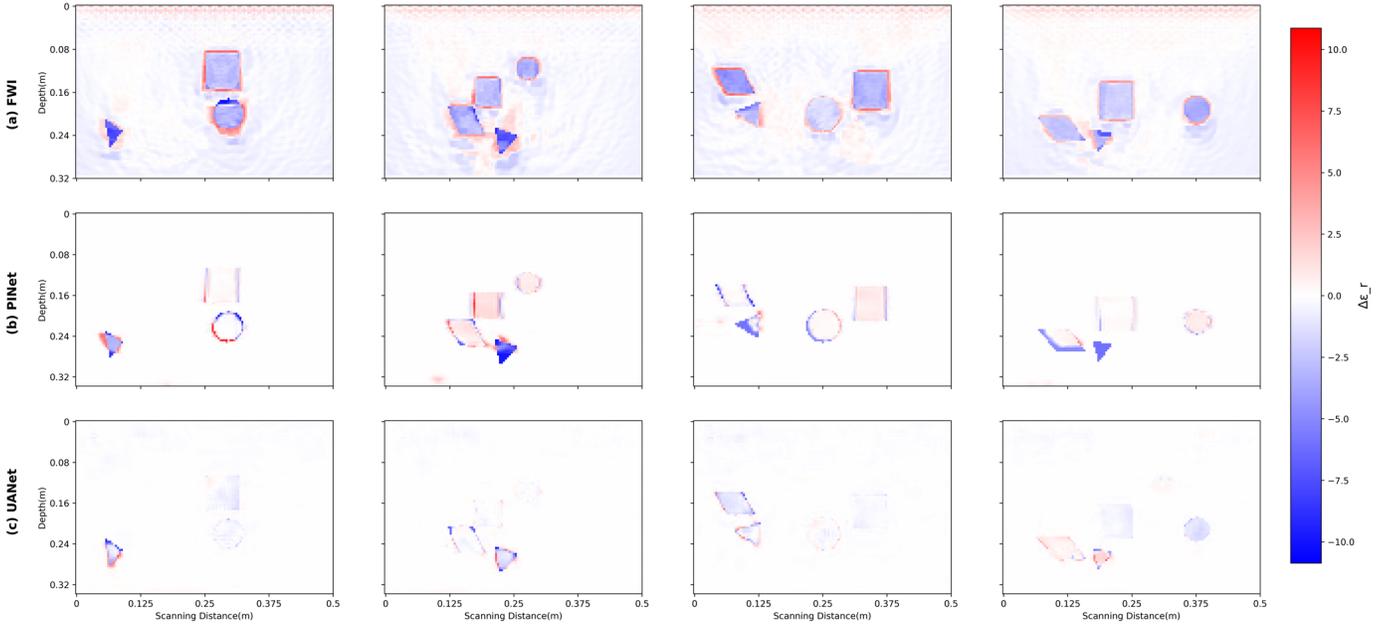

Fig. 9. Difference maps between the reconstructed and ground-truth permittivity distributions using (a) FWI, (b) PINet, and (c) UA-Net.

TABLE II
COMPARISON OF AVERAGE MRT(s), SSIM, AND PSNR (dB) FOR FOUR TEST SAMPLES

| Method | MRT | SSIM | PSNR |
| --- | --- | --- | --- |
| FWI | 2623.45s | 0.6968 | 20.58 |
| PINet | 0.42s | 0.9464 | 24.23 |
| UA-Net | 0.67s | 0.9742 | 30.55 |

SSIM and PSNR of UA-Net are significantly improved, indicating that the permittivity distribution reconstructed by UA-Net is closer to ground truth. There is also a large disparity in computational cost between classical FWI and UA-Net. The MRT of classical FWI is about 43 minutes (150 iterations). In contrast, once UA-Net is well-trained, the inference phase takes only 0.67 seconds. Compared to the purely data-driven PINet, the PSNR and SSIM of UA-Net are also improved, which demonstrates that the network's inverse capability can be enhanced by incorporating domain knowledge. Although the MRT of UA-Net is higher than that of PINet, this additional computational cost is acceptable because UA-Net reduces the quantitative demand for training data, which is especially important in scenarios where labeled data acquisition is challenging.

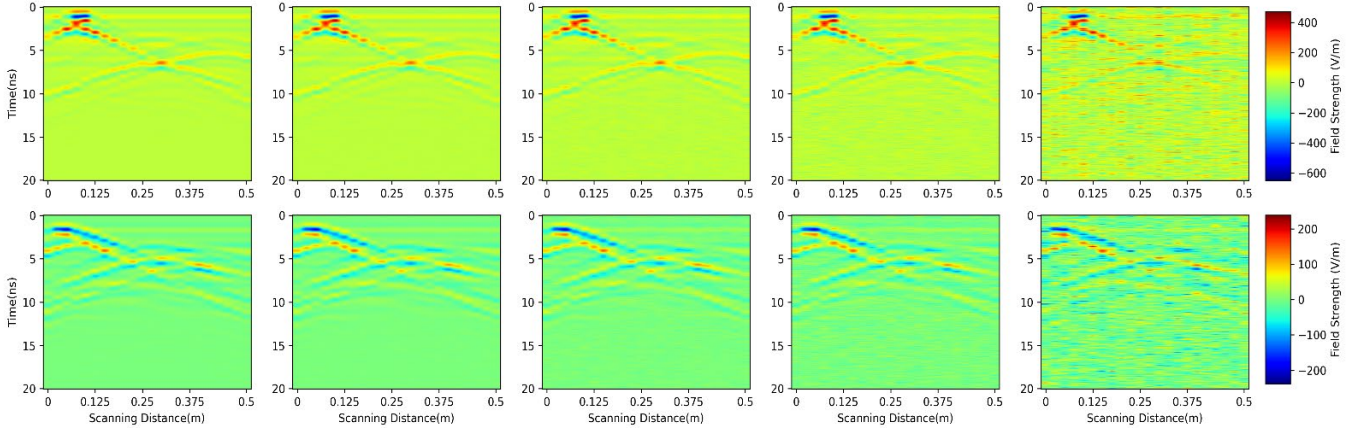

Fig. 10. Noise-added B-scans. The first column shows the original direct-wave removal B-scans. The second, third, fourth, and fifth columns show the B-scans with SNR values of 20 dB, 10 dB, 5 dB, and -5 dB, respectively.

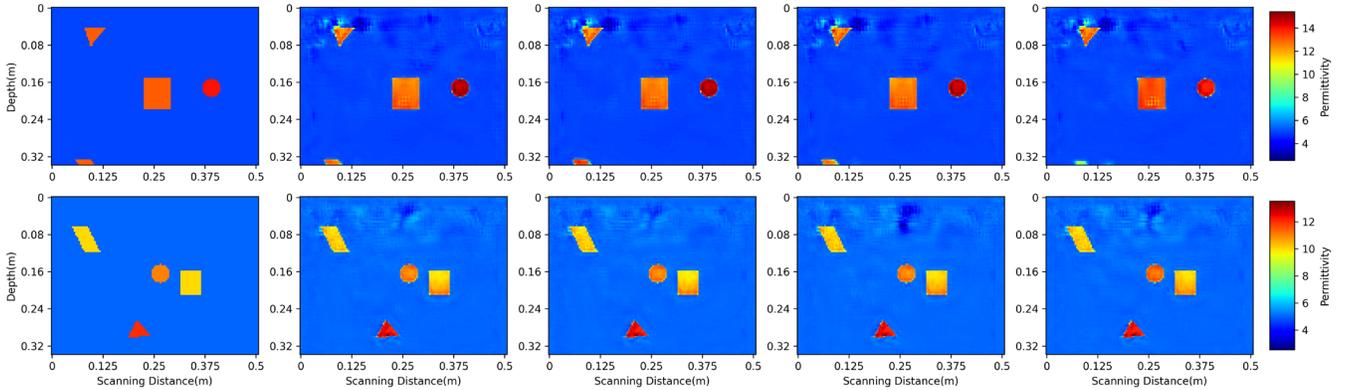

Fig. 11. Inversion results of noise-added B-scans. The first column shows the ground truth. The second, third, fourth, and fifth columns show the inversion results with SNR values of 20, 10, 5, and -5, respectively.

*C. Experiment III: Noise Robustness Tests*

To verify the noise robustness of UA-Net, the Gaussian white noise is added to the direct-wave removal test B-scan generated in Experiment I. Noise-added B-scans are applied to the pre-trained UA-Net without additional training. Fig. 10 presents two test samples with specified signal-to-noise ratio (SNR). The first column shows the original B-scan without noise, while the second, third, fourth, and fifth columns show the noise-added B-scans with SNR of 20 dB, 10 dB, 5 dB, and -5 dB, respectively.

Fig. 11 shows the inversion results of noise-added B-scans. As the SNR decreases, the deep structure of the first sample is poorly reconstructed. However, the shallow structure is still accurately reconstructed, primarily due to the introduction of the learnable regularizer. The regularizer suppresses noise of the object area, thereby enhancing the reconstruction capability of UA-Net in the complex background (with noise).

*D. Experiment IV: Generalizability Experiments*

To test the generalization ability of the proposed UA-Net in reconstructing permittivity distributions in different environments, the performance of UA-Net and the data-driven PINet is evaluated using four layered models. Since the layered models differ from those in the training dataset, transfer learning is introduced to adjust the pre-trained network. The DL-based GPR solver used in UA-Net is fine-tuned with a dataset containing 4,000 samples. To ensure a fair comparison, UA-Net and PINet are both fine-tuned using dataset with 300 samples and a learning rate of 0.0001 during inversion. After 50 epochs of training, the predictions of the two networks are presented in Fig. 12. Fig. 12 presents the inversion results of the PINet and UA-Net for the four layering models. In Fig. 12 Case 1 and Case 2, PINet predicts the correct background layering, but the objects with low permittivity suffer from significant distortion. In Fig. 12 Case 3, PINet fails to predict the background layering, and the object size and permittivity show significant differences. In Fig. 12 Case 4, PINet fails to reconstruct the permittivity distribution of multiple objects. These results show that fine-tuning a purely data-driven network with a small dataset does not yield desirable results. In contrast, as shown in Fig. 12 (c), the predictions of the UA-Net match well with the ground truths. By fine-tuning the well-trained network with a small dataset, UA-Net can still reconstruct the permittivity distribution with high accuracy.

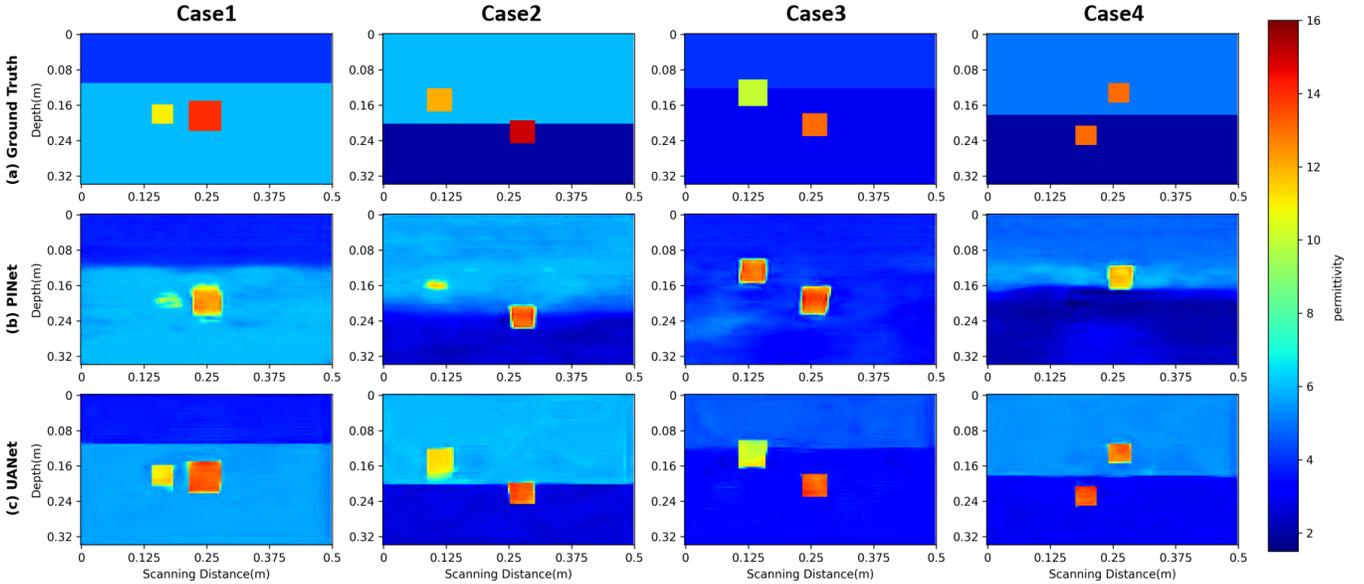

Fig. 12. Generalizability testing results. (a) Ground truths of the permittivity; (b) reconstructed permittivity distributions using PINet; (c) predicted permittivity distributions using UA-Net.

TABLE III
PSNR (dB) AND SSIM FOR FOUR TEST SAMPLES

| Metrics | Method | T1 | T2 | T3 | T4 |
|---|---|---|---|---|---|
| PSNR | PINet | 23.75 | 21.29 | 22.05 | 19.29 |
|  | UA-Net | **29.42** | **25.51** | **32.55** | **26.60** |
| SSIM | PINet | 0.8597 | 0.8023 | 0.8476 | 0.7753 |
|  | UA-Net | **0.9643** | **0.9203** | **0.9360** | **0.9418** |

The comparison with the two networks on the evaluation metrics is provided in TABLE III. The SSIM and PSNR of the reconstructed permittivity distributions using UA-Net are much better than PINet. Moreover, the performance metrics of UA-Net closely match those of the pre-trained model, while the SSIM of PINet shows a significant decrease compared to the pre-trained model. This indicates that the predictions of UA-Net maintain a strong similarity to the ground truth in new environments.

*E. Experiment V: Real Measurement Data Experiments*

The field dataset provided by Dai et al. [18] was employed to further evaluate the capability of UA-Net in reconstructing subsurface permittivity distributions. Real B-scans were acquired on uneven sandy ground using a commercial GSSI Utility Scan Pro ground-penetrating radar system equipped with a 400 MHz antenna. Each B-scan acquired along a one-meter scanning trace and consisted of 88 A-scans, each containing 512 sampling points. The data acquisition time window was set to 20 ns. The buried objects comprised five wooden objects with varying shapes, sizes, and relative permittivity. Noise was removed from the original B-scan by subtracting data collected in a region without buried objects, providing the denoised Bscans. A total of 196 measurement sets were used in this study, including noisy B-scans, denoised B-scans, and the corresponding subsurface permittivity distributions.

Since the simulation data in our study differed from the actual measurement scenario, UA-Net was first pre-trained using 18,000 simulated datasets provided by Dai et al. [18], followed by fine-tuning on the 196 measured datasets via a transfer learning strategy. Both simulated and measured datasets were split into training, validation, and test sets in an 8:1:1 ratio. As shown in Fig. 13, the proposed method accurately predicted the permittivity of buried objects in sandy soil. In Case 1, UA-Net reconstructed overall location of the rectangular object, though the boundaries were somewhat blurred. In Case 2, the circular object was well localized, and the reconstructed permittivity closely matched the ground truth. In Case 3, both rectangular and circular objects were reconstructed with accurate overall positions, demonstrating that the method maintains strong inverse capability in multi-object scenarios. However, some deviations were observed in Case 3: the permittivity along the edges of the circular object was slightly underestimated, which may be attributed to the five-layer iterative structure of UA-Net, as objects with high permittivity may require deeper iterations to improve inversion accuracy. The rectangular object was not fully continuous, and its lateral extent was slightly smaller than the ground truth. Compared to the circular object, the rectangular shape was less accurately reconstructed, possibly because the pre-training data included more circular objects but lacked elongated rectangular samples, leading the network to favor more uniform shapes during reconstruction.

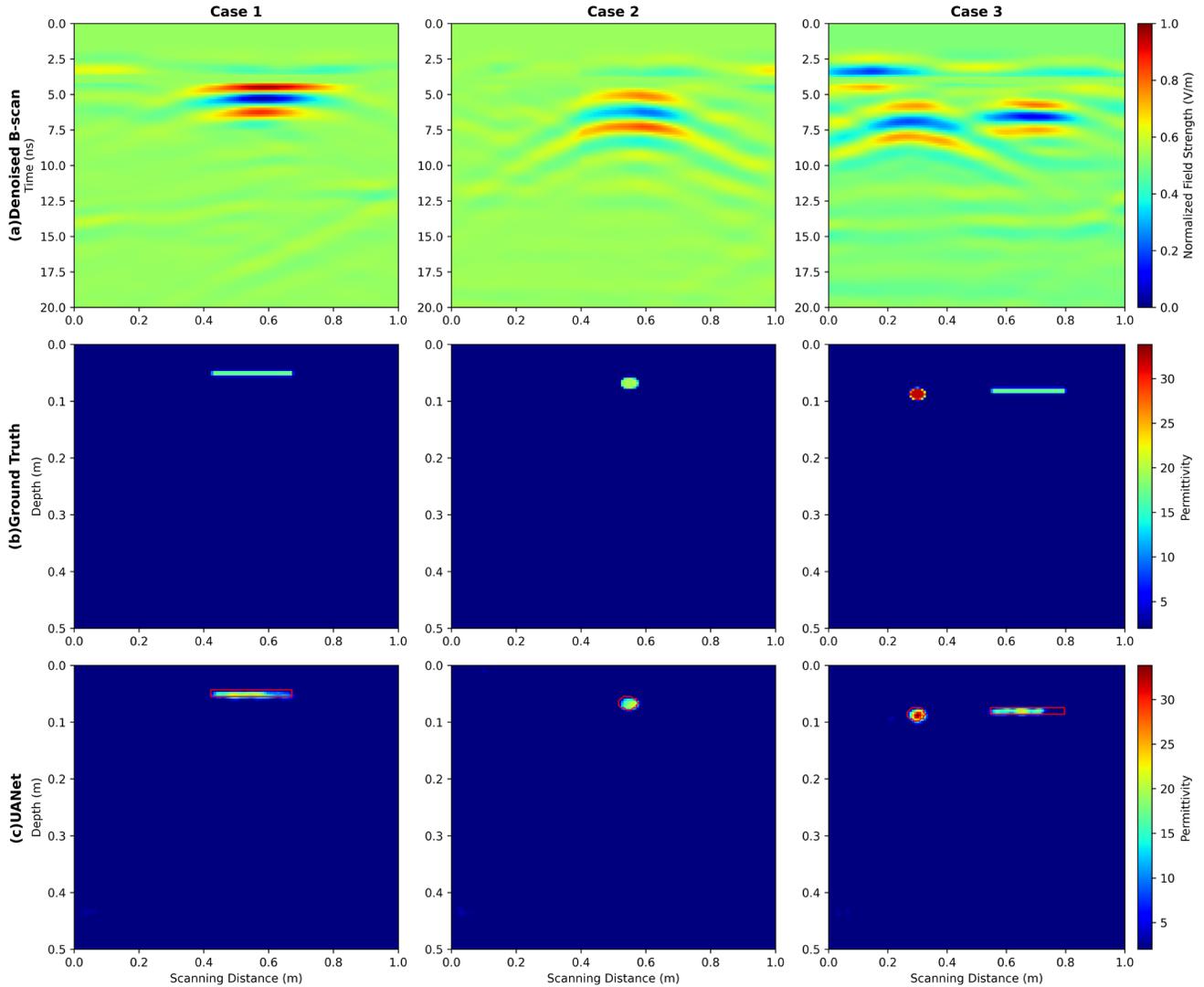

Fig. 13. Comparison of permittivity distributions reconstructed by UA-Net with the ground truth for three real-world test cases.

IV. DISCUSSION

In this study, we propose a fully neural network-based structure, which is developed based on a DL-based GPR solver and iterative unfolding strategy. To validate the effectiveness of the proposed method, we compare it with classical FWI and a representative data-driven network, PINet. Experimental results show that the proposed UA-Net outperforms classical FWI and PINet. In this section, we discuss the advantages, limitations and future directions of the proposed UA-Net.

For classical FWI, a smoothed initial permittivity model and multiscale strategy are employed to mitigate getting trapped in local minima. Nevertheless, the PSNR and SSIM in TABLE I demonstrate that the reconstruction results of UA-Net are superior to classic FWI. Additionally, UA-Net has an advantage in computational efficiency, with an MRT of 0.67 compared to 2623.45 for the classic FWI. For network training, a total of 20 000 samples are used to train PINet, while UA-Net produces superior results with only 1000 samples in the inversion phase. This demonstrates that UA-Net is suitable for ground-penetrating radar scenarios with sparse labeled samples. In the inference phase, UA-Net requires more time than PINet, but the computational overhead remains acceptable. Furthermore, in the out-of-distribution generalization experiments, PINet exhibits different degrees of distortion, while UA-Net can yield superior predictions. The experiments indicate that the regularizer learned by UA-Net improves the reconstruction accuracy of permittivity while also being robust to different scenarios. Finally, the effectiveness of UA-Net is further validated through real data experiments, confirming its practical applicability.

To further advance the current methodology, we are developing an extension for dual-parameter inversion, where a parallel unfolding network architecture enables simultaneous updates of both permittivity and conductivity. Although the proposed approach has been validated on both synthetic and real GPR data, it is worth noting that actual measurements are often affected by factors such as noise, heterogeneous underground materials, and missing data, which may differ significantly from the synthetic training data. Therefore, enhancing the realism of synthetic datasets remains an important direction for future research.

## V. Conclusion

In this study, we propose a fully neural network-based model-driven solution for GPR FWI, named UA-Net. This approach transforms the traditional iterative algorithm into a multi-stage network and integrates a DL-based GPR solver to accelerate the computation of data misfit. Notably, the nonlocal regularization module trained alongside other network parameters, contributes to improve reconstruction quality. The effectiveness of UA-Net is validated on both synthetic and real GPR data. Future work will focus on extending the framework to dual-parameter inversion, enabling the simultaneous recovery of permittivity and conductivity for more comprehensive subsurface characterization.

## Appendix

To improve the reconstruction performance, the matrix $D$ in (7) is defined as a nonlinear transformation function, denoted as $F(\cdot)$:

$$z_k = \arg\min_z \frac{\gamma_k}{2} \| \varepsilon_k + \frac{1}{\gamma_k}\alpha_{k-1} - z_{k-1} \|_2^2 + \lambda_k \| F(z_{k-1}) \|_1 \quad (A1)$$

Note that $z_k$ is the immediate reconstruction result of $\varepsilon_k$ at the k-th iteration. According to Theorem 1 of Zhang and Ghanem [39], assuming that $\varepsilon_k + 1/\gamma_k \alpha_{k-1}$ and $F(\varepsilon_k + 1/\gamma_k \alpha_{k-1})$ are the average values of $z_{k-1}$ and $F(z_{k-1})$, respectively, we can obtain:

$$\| F_k\left(\varepsilon_k + \frac{1}{\gamma_k}\alpha_{k-1}\right) - F_k(z_{k-1}) \|_2^2$$
$$\approx \tau \| \varepsilon_k + \frac{1}{\gamma_k}\alpha_{k-1} - z_{k-1} \|_2^2 \quad (A2)$$

where $\tau$ is a scalar only related to $F(\cdot)$.

By incorporating (A2) into (A1), it yields the following optimization:

$$z_k = \arg\min_z \frac{1}{2} \| F_k\left(\varepsilon_k + \frac{1}{\gamma_k}\alpha_{k-1}\right) - F_k(z_{k-1}) \|_2^2 + \theta_k \| F(z_{k-1}) \|_1 \quad (A3)$$

where $\theta_k = \lambda_k \tau / \gamma_k$.

Therefore, we obtain a closed form version of $F(z_{k-1})$:

$$F_k(z_k) = soft\left(F_k\left(\varepsilon_k + \frac{1}{\gamma_k}\alpha_{k-1}\right), \theta_k\right) \quad (A4)$$

We design the transform $G(\cdot)$, which has a symmetric structure with $F(\cdot)$, $z_k$ can be efficiently computed as follows:

$$z_k = G_k\left(soft\left(F_k\left(\varepsilon_k + \frac{1}{\gamma_k}\alpha_{k-1}\right), \theta_k\right)\right) \quad (A5)$$